\documentstyle[12pt,epsfig,a4]{article} 
\newcommand{\vs}{\vspace{-0.25cm}}
\begin{document} 

\begin{center}
{\Large{\bf Cross sections for low-energy $\pi^-\gamma$ reactions}}  
\bigskip

N. Kaiser and J.M. Friedrich\\
\medskip
{\small Physik-Department, Technische Universit\"{a}t M\"{u}nchen,
    D-85747 Garching, Germany}
\end{center}
\medskip
\begin{abstract}
We review the cross sections for low-energy $\pi^- \gamma$ reactions in the 
framework of chiral perturbation theory. Charged pion Compton scattering, 
$\pi^- \gamma\to \pi^-\gamma$, is considered up to one-loop order where the 
pion's internal structure enters through the difference of the electric and 
magnetic pion polarizability, $\alpha_\pi - \beta_\pi$. The ongoing COMPASS 
experiment aims at measuring this important structure constant with high 
statistics using the Primakoff effect. In the same way, the reaction $\pi^- 
\gamma \to \pi^- \pi^0$ serves as a test of the QCD chiral anomaly (via the 
$\gamma 3\pi$-coupling constant $F_{\gamma3\pi}$). Furthermore, we calculate the 
total cross sections for neutral and charged pion-pair production, $\pi^- 
\gamma \to \pi^- \pi^0\pi^0$ and $\pi^- \gamma \to \pi^-\pi^+\pi^-$, which are 
governed by the chiral $\pi\pi$-interaction. Finally, we investigate the 
radiative (correction) process $\pi^- \gamma \to \pi^- \gamma \gamma$ and 
calculate the corresponding two-photon mass spectrum. This information will be 
useful for analyzing the $\pi^-$ Primakoff scattering events with photons in
the final state.     
\end{abstract}

\bigskip

\section{Introduction and summary}
The pions $(\pi^+,\pi^0,\pi^-)$ are the Goldstone bosons of spontaneous chiral 
symmetry breaking in QCD: $SU(2)_L\times SU(2)_R \to SU(2)_V$. Their low-energy 
dynamics can therefore be calculated systematically (and accurately) with 
chiral perturbation theory in form of a loop-expansion based on an effective 
chiral Lagrangian. The very accurate two-loop prediction \cite{cola} for the 
isospin-zero S-wave $\pi\pi$-scattering length $a_0^0 = (0.220 \pm 0.005)
m_\pi^{-1}$ has recently been confirmed in the E865 \cite{bnl} and NA48/2
\cite{batley} experiments 
by analyzing the $\pi^+\pi^-$ invariant mass distribution of the rare kaon
decay mode $K^+ \to \pi^+\pi^-e^+\nu_e$. One particular implication of that
good agreement between theory and experiment is that the quark condensate
$\langle 0|\bar q q |0\rangle $ constitutes the leading order parameter
\cite{condensate} of spontaneous chiral symmetry breaking (considering the
two-flavor sector of QCD). As a consequence, one can assert that more than
$90\%$ of the (squared) pion mass, $m_\pi=139.57\,$MeV/$c^2$, must stem from the
term linear in the light quark mass multiplied with the quark condensate 
$\langle 0|\bar q q |0\rangle$ \cite{condensate}. Furthermore, the DIRAC 
experiment \cite{dirac} at CERN has been proposed to determine the difference 
of the isospin-zero and isospin-two S-wave $\pi\pi$-scattering lengths $a_0^0
-a_0^2$ by measuring the life time ($\tau \simeq 3\,$fs) of pionium (i.e. 
$\pi^+\pi^-$ bound electromagnetically and decaying into $\pi^0\pi^0$). In the 
meantime the NA48/2 collaboration \cite{cusp} at CERN has accumulated very 
high statistics for the charged kaon decay modes $K^\pm \to \pi^\pm \pi^0\pi^0$, 
which allowed  them to extract the value $a_0^0-a_0^2=(0.268\pm 0.010)m_\pi^{-1}$ 
for the $\pi\pi$-scattering length difference from the cusp effect in the 
$\pi^0\pi^0$ mass spectrum at the $\pi^+\pi^-$ threshold. This experimental
result is again in very good agreement with the two-loop prediction
$a_0^0-a_0^2 = (0.265 \pm 0.004)m_\pi^{-1}$ of chiral perturbation theory 
\cite{cola}. For a discussion of isospin breaking corrections which have to be 
included in a meaningful comparison between theory and experiment, see also 
ref.\cite{gasserisobr}. Clearly, these remarkable confirmations give
confidence that chiral perturbation theory is the correct framework to
calculate reliably and accurately the strong interaction dynamics of the pions
at low energies.  

Electromagnetic processes offer further possibilities to probe the internal
structure of the pion. Whereas the charge (or vector) form factor of the pion, 
$F_\pi(t)$, is largely dominated by the low-lying $\rho(770)$-resonance, pion 
Compton scattering $\pi^- \gamma\to \pi^-\gamma$ allows one to extract the
electric and magnetic polarizabilities of the (charged) pion. In a classical
picture these polarizabilities characterize the deformation response
(i.e. induced dipole moments) of a composite system in external electric and
magnetic fields. In the proper quantum field theoretical  formulation the
electric and magnetic polarizabilities, $\alpha_\pi$ and $\beta_\pi $, are 
defined as expansion coefficients of the Compton scattering amplitudes at 
threshold. Since pion targets are not directly available, real pion Compton 
scattering has been approached using different artifices, such as high-energy 
pion-nucleus bremsstrahlung $\pi^- Z \to \pi^- Z \gamma$ \cite{serpukov}, 
radiative pion production off the proton $\gamma p \to \gamma\pi^+ n$ 
\cite{mainz}, and the crossed channel two-photon reaction $\gamma \gamma \to 
\pi^+ \pi^-$ as embedded in the $e^+e^-  \to e^+e^- \pi^+ \pi^-$ process 
\cite{mark}. The corresponding results for the electric and magnetic pion 
polarizabilities scatter substantially and they have also large uncertainties. 
On the other hand, chiral perturbation theory at two-loop order gives the firm 
prediction $\alpha_\pi- \beta_\pi =(5.7\pm1.0)\cdot 10^{-4}\,$fm$^3$ \cite{gasser} 
for the polarizability difference. It is however in conflict with the existing 
experimental results from Serpukhov $\alpha_\pi-\beta_\pi= (15.6\pm 7.8)\cdot 
10^{-4}\,$fm$^3$ \cite{serpukov} and MAMI $\alpha_\pi-\beta_\pi= (11.6\pm
3.4)\cdot 10^{-4}\,$fm$^3$ \cite{mainz} which amount to values more than twice as
large. The result $\alpha_\pi-\beta_\pi=(4.4\pm  3.2)\cdot 10^{-4}\, $fm$^3$
\cite{doho} extracted from the MARK II data \cite{mark} is consistent with the 
chiral prediction, but the corresponding low-energy cross sections for $\gamma
\gamma \to \pi^+ \pi^-$ are (within their errorbars) rather insensitive to even 
sizeable changes of the pion polarizabilities \cite{filkov}. We also note that 
radiative pion photoproduction $\gamma p \to \gamma\pi^+ n$ has recently  been
considered in the framework of heavy baryon chiral perturbation theory 
\cite{kao}. It has been argued in ref.\cite{kao} that additional contributions 
from $\gamma\gamma \pi NN$ vertices in the chiral Lagrangian ${\cal L}_{\pi  
N}^{(3)}$ may have effects on the cross section which are comparable to those 
of the pion polarizabilities.

In that situation, it is very promising that the ongoing COMPASS experiment
\cite{compass} at CERN aims at measuring the pion polarizabilities, 
$\alpha_\pi$ and $\beta_\pi$, with high statistics using the Primakoff 
effect. The scattering of high-energy negative pions in the Coulomb field of a 
heavy nucleus (of charge $Z$) gives access to cross sections for $\pi^-\gamma$ 
reactions through the equivalent photon method:
\begin{equation} {d \sigma \over ds dt dQ^2} = {Z^2 \alpha\over \pi(s-m_\pi^2)}
\, {Q^2-Q_{min}^2 \over Q^4}\,\, {d \sigma_{\pi^- \gamma} \over dt}\,, 
\qquad Q_{min} = {s-m_\pi^2 \over 2E_{\rm beam}}\,. \end{equation}    
Here, $\sqrt{s}>m_\pi$ is the $\pi^-\gamma$ center-of-mass energy and $t<0$ the 
squared invariant momentum transfer between the initial and final state
pion. $Q$ denotes the momentum transferred by the virtual photon to the heavy
nucleus, and one aims at isolating the Coulomb peak $Q\to 0$ from the strong  
interaction background. Due to the small range of $Q \approx 0$ the nuclear
charge form factor can be approximated by $F_Z(Q^2) \simeq Z$. A first 
pilot run performed at COMPASS has already accumulated statistics for 
Primakoff events comparable to previous experiments of this type 
\cite{serpukov,antipov}. For a recent discussion of possible problems (nuclear 
scattering, kinematical limitations etc.) associated with isolating the 
Coulomb peak, see refs.\cite{faeldt,faeldtsigma}. 
 
The purpose of the present paper is to review the cross sections for various
low-energy $\pi^-\gamma$ reactions in the framework of chiral perturbation
theory. We treat pion Compton scattering, $\pi^- \gamma\to \pi^-\gamma$, up to 
one-loop order, where the pion's internal structure appears for the first time 
in form of the difference of the electric and magnetic polarizability,
$\alpha_\pi -\beta_\pi$. In the view the expected accuracy of the COMPASS 
experiment, and the knowledge that two-loop corrections \cite{gasser} are
relatively small, this approximation should be sufficient. Then, we consider
(at one-loop order) the reaction $\pi^- \gamma \to\pi^- \pi^0$ which is also
under study in the COMPASS experiment. It serves as a test of the QCD chiral
anomaly via measuring the $\gamma 3\pi$-coupling constant $F_{\gamma3\pi}$. As
the pertinent one-loop amplitudes are available from the literature, we
restrict here ourselves to a concise but explicit presentation of the relevant
cross section formulas. Actually new is our calculation of total cross
sections for neutral and charged pion-pair production, $\pi^- \gamma \to \pi^-
\pi^0\pi^0$ and $\pi^- \gamma \to \pi^- \pi^+\pi^-$, which interestingly are
determined by the chiral $\pi\pi$-interaction. With the larger data sample
planned in future COMPASS runs sufficient statistics will become available for
these reactions. Finally, we investigate the radiative correction process
$\pi^- \gamma \to \pi^- \gamma \gamma$ and calculate the corresponding
two-photon mass spectrum. This information will be helpful for analyzing the
$\pi^-$ Primakoff scattering events with photons in the final state.     
\section{Charged pion Compton scattering}
We start with the process of Compton scattering off a negatively charged pion:
$\pi^-(p_1)+\gamma(k_1,\epsilon_1)\to \pi^-(p_2)+\gamma(k_2,\epsilon_2)$. The 
corresponding T-matrix in the center-of-mass frame has (in Coulomb gauge 
$\epsilon_{1,2}^0=0$) the form:
\begin{equation} 
T_{\pi \gamma} = 8\pi \alpha \Big\{ - \vec \epsilon_1
\cdot\vec \epsilon_2 \, A(s,t) +  \vec \epsilon_1 \cdot \vec k_2\,  \vec 
\epsilon_2 \cdot \vec k_1 \, {2\over t} \Big[  A(s,t)+ B(s,t) \Big] \Big\}\,, 
\end{equation}
with $\alpha=1/137.036$ the fine-structure constant, and $s=(p_1+k_1)^2>m_\pi^2$ 
and  $t=(k_1-k_2)^2<0$ the independent Mandelstam variables. Performing the
sums over transversal polarizations and applying flux and appropriate phase
space factors, the resulting  differential cross section reads: 
\begin{equation} {d \sigma \over d \Omega_{\rm cm}} = {\alpha^2 \over
2s} \Big\{ |A(s,t)|^2 +|A(s,t)+(1+z)B(s,t)|^2 \Big\} \,, \end{equation}
with $t = (s-m_\pi^2)^2(z-1)/2s$ where $z = \cos \theta_{\rm cm}=\hat k_1
\cdot \hat k_2$ is the cosine of the cms scattering angle. The decomposition
in eq.(2) into two invariant amplitudes  $A(s,t)$ and $B(s,t)$ has been done
with hindsight to an expression as simple as possible for the (unpolarized) 
differential cross section. 

The amplitudes at tree level coincide with the ones from scalar quantum 
electrodynamics:   

\begin{equation} A(s,t)^{(\rm tree)} = 1, \qquad\qquad  B(s,t)^{(\rm tree)} =
{s-m_\pi^2 \over m_\pi^2-s-t}\,. \end{equation}
Note that the contribution of the $s$-channel pole diagram vanishes in Coulomb
gauge since the coupling of the initial state photon becomes zero in that 
gauge, $\epsilon_1\cdot(2p_1+k_1)=0$. The one-pion loop diagrams of chiral 
perturbation theory generate, after renormalization of the pion mass $m_\pi$,
the following (finite) contribution to the Compton amplitude $A(s,t)$
\cite{gasser,buergi}:   
\begin{equation} A(s,t)^{(\rm loop)} = -{1\over (4\pi f_\pi)^2} \Bigg\{ {t 
\over 2} + 2m_\pi^2 \ln^2 {\sqrt{4m_\pi^2-t}+\sqrt{-t} \over 2m_\pi} \Bigg\}\,,
\end{equation}
with $f_\pi = 92.4\,$MeV the pion decay constant. Heuristically, it can be 
interpreted as the (leading) correction arising from photon scattering off the 
''pion cloud around the pion''. The internal structure of the pion enters
through its electric and magnetic polarizabilities, which obey at the one-loop
order the constraint $\alpha_\pi+ \beta_\pi=0$ \cite{holstein}. The pertinent 
$\gamma \gamma \pi \pi$ contact vertices from the chiral Lagrangian ${\cal L}_{
\pi\pi}^{(4)}$ give rise to the contribution \cite{gasser,buergi}:  
\begin{equation}A(s,t)^{(\rm pola)}=-{\beta_\pi m_\pi t \over    2\alpha}\,, 
\qquad\qquad  \alpha_\pi -\beta_\pi ={\alpha (\bar  l_6-\bar l_5)
\over 24 \pi^2 f_\pi^2 m_\pi}\,, \end{equation} 
to the Compton amplitude $A(s,t)$. The relevant combination of low-energy 
constants $\bar l_6 - \bar l_5$ can be extracted from the axialvector-to-vector 
form factor ratio $h_A/h_V = 0.443 \pm 0.015 =(\bar l_6 - \bar l_5)/6+{\cal O}
(m_\pi^2)$ measured in the PIBETA experiment \cite{frlez} at PSI via the 
radiative pion decay $\pi^+\to e^+\nu_e\gamma$. The two-loop analysis of
ref.\cite{geng} yields the value $ \bar l_6 - \bar l_5 =3.0\pm 0.3$ implying
the pion polarizability difference: $\alpha_\pi - \beta_\pi \simeq 6.0 \cdot
10^{-4}\,$fm$^3$. It should be stressed that the current-algebra relation 
$\alpha_\pi - \beta_\pi =\alpha h_A/(4\pi^2 h_V f_\pi^2 m_\pi)+ {\cal  O}(m_\pi)$ 
constitutes a low-energy theorem \cite{terentev} which must hold (to leading 
order) in any chiral invariant theory. From this point of view, the  
parametrization of $\alpha_\pi - \beta_\pi$ in terms of a phenomenological 
$\sigma$-meson exchange, as chosen in ref.\cite{faeldtsigma}, is problematic 
since it ignores that profound theoretical constraint. Moreover, as
stressed in ref.\cite{holstein} the outcome $\bar l_6 - \bar l_5=-2$ of the
chiral invariant linear $\sigma$-model is completely ruled out by experiment,
because it leads to an axialvector-to-vector form factor ratio $h_A/h_V$ of the
wrong sign. As a matter of fact, the actual physics behind the low-energy
constant $\bar l_6 - \bar l_5$ is the excitation of vector and axialvector
meson resonances \cite{holstein}. In the same way the existing experimental 
determinations of $\alpha_\pi - \beta_\pi\simeq 12\cdot 10^{-4}\,$fm$^3$
\cite{serpukov,mainz} give reason to doubts since they violate the chiral 
low-energy theorem notably by a factor $2$. In order to clarify the situation
on the theoretical side, a complete two-loop calculation of pion
polarizabilities has been performed in ref.\cite{gasser,buergi}, grinding  out
the analytical result:  
\begin{eqnarray}\alpha_\pi -\beta_\pi &=& {\alpha (\bar 
l_6-\bar l_5) \over 24\pi^2 f_\pi^2 m_\pi} + {\alpha m_\pi\over (4\pi f_\pi)^4}
\bigg\{c^r+{8\over 3}\bigg( \bar l_2-  \bar l_1+ \bar l_5-\bar l_6+{65\over 12} 
\bigg)\ln{m_\pi \over m_\rho} \nonumber \\ && +{4\over 9}(\bar l_1+ \bar l_2)-{ 
\bar l_3\over 3} +{4 \bar l_4\over 3}(\bar l_6-\bar l_5) -{187 \over 81} +
\bigg( {53 \pi^2 \over 48} -{41 \over 324 }\bigg) \bigg\}\,, \end{eqnarray}
expressed in terms of the chiral low-energy constants $\bar l_1 = -0.4\pm 0.6$, 
$\bar l_2 = 4.3\pm 0.1$, $\bar l_3 = 2.9\pm 2.4$, $\bar l_4 = 4.4\pm 0.2$. The
last term proportional to $53 \pi^2/48 -41/324$ stems from the nonfactorizable
acnode diagrams.  The additional counterterm $c^r$ has been estimated via 
resonance saturation as $c^r \simeq 0$ \cite{gasser} and we have set the scale 
in the chiral logarithm $\ln(m_\pi/m_\rho)$ equal to the $\rho$-meson mass 
$m_\rho = 770\,$MeV/$c^2$. Taking into account various theoretical uncertainties,
the two-loop prediction amounts to: $\alpha_\pi -\beta_\pi =(5.7\pm 1.0)\cdot 
10^{-4}\,$fm$^3$ \cite{gasser}. As this value is fully compatible with the 
current-algebra result, it is now assured that there are no significant 
corrections to the chiral low-energy theorem for the pion polarizability 
difference $\alpha_\pi- \beta_\pi$. For a more detailed discussion of these 
issues, see ref.\cite{gasser}. The non-vanishing two-loop prediction for the 
pion polarizability sum, $\alpha_\pi+ \beta_\pi= 0.16 \cdot 10^{-4}\,$fm$^3$ 
\cite{gasser}, is in fact consistent with results from dispersion sum  rules 
\cite{filkov} but presumably too small to cause an observable effect in 
pion Compton scattering. Moreover, as indicated e.g. by Fig.\,6 in 
ref.\cite{gasser} the one-loop approximation should be sufficient for low 
energies $\sqrt{s}<4m_\pi$.

\begin{figure}[htb]
\begin{center}
\includegraphics[scale=.55,clip]{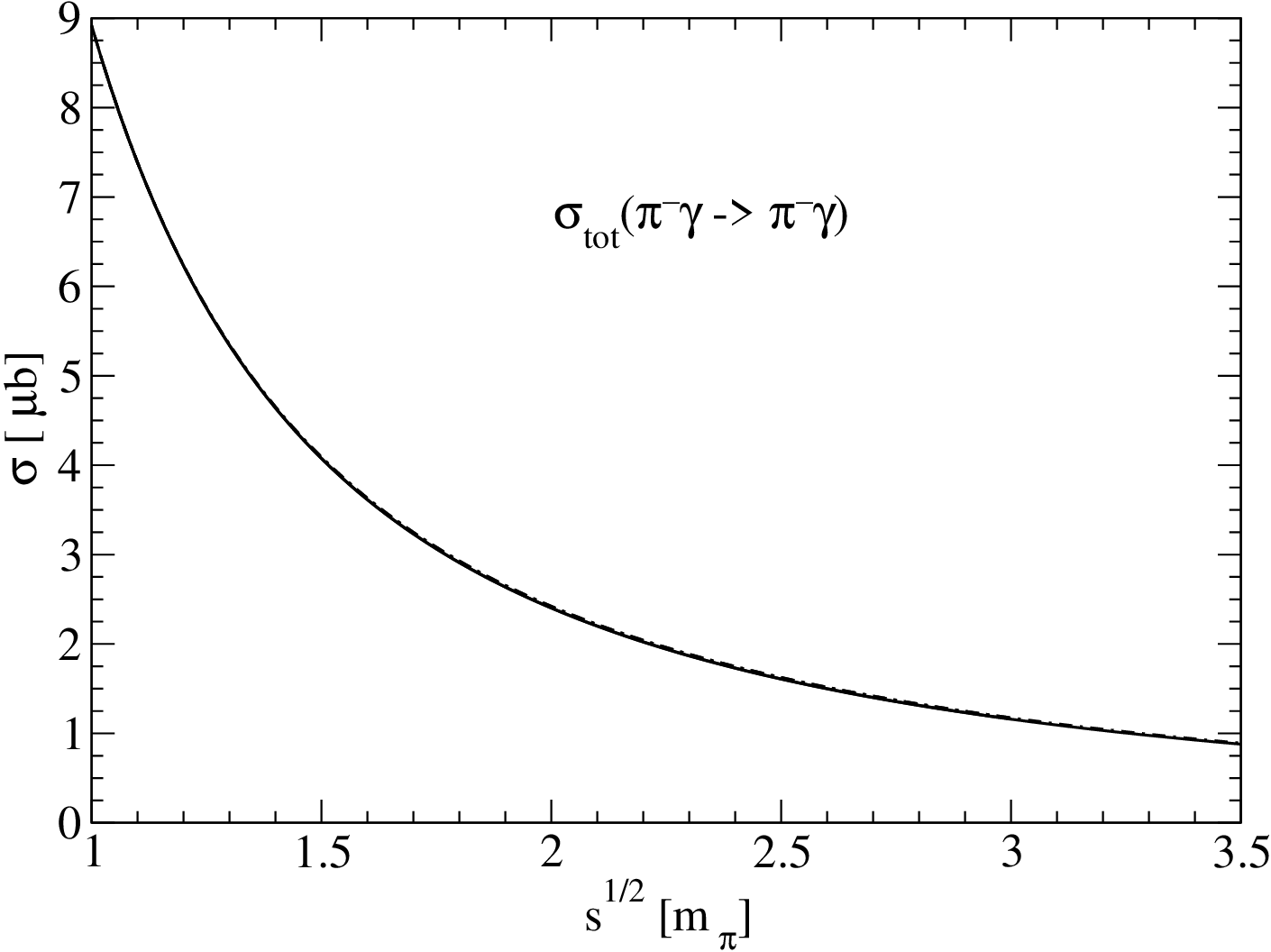}
\end{center}
\vspace{-.8cm}
\caption{Total elastic cross section for charged pion Compton scattering. 
Effects from the pion's low-energy structure (pion-loops and polarizability
difference $\alpha_\pi -\beta_\pi$) contribute at the level of $1\%$.}
\end{figure}
\begin{figure}
\begin{center}
\includegraphics[scale=.55,clip]{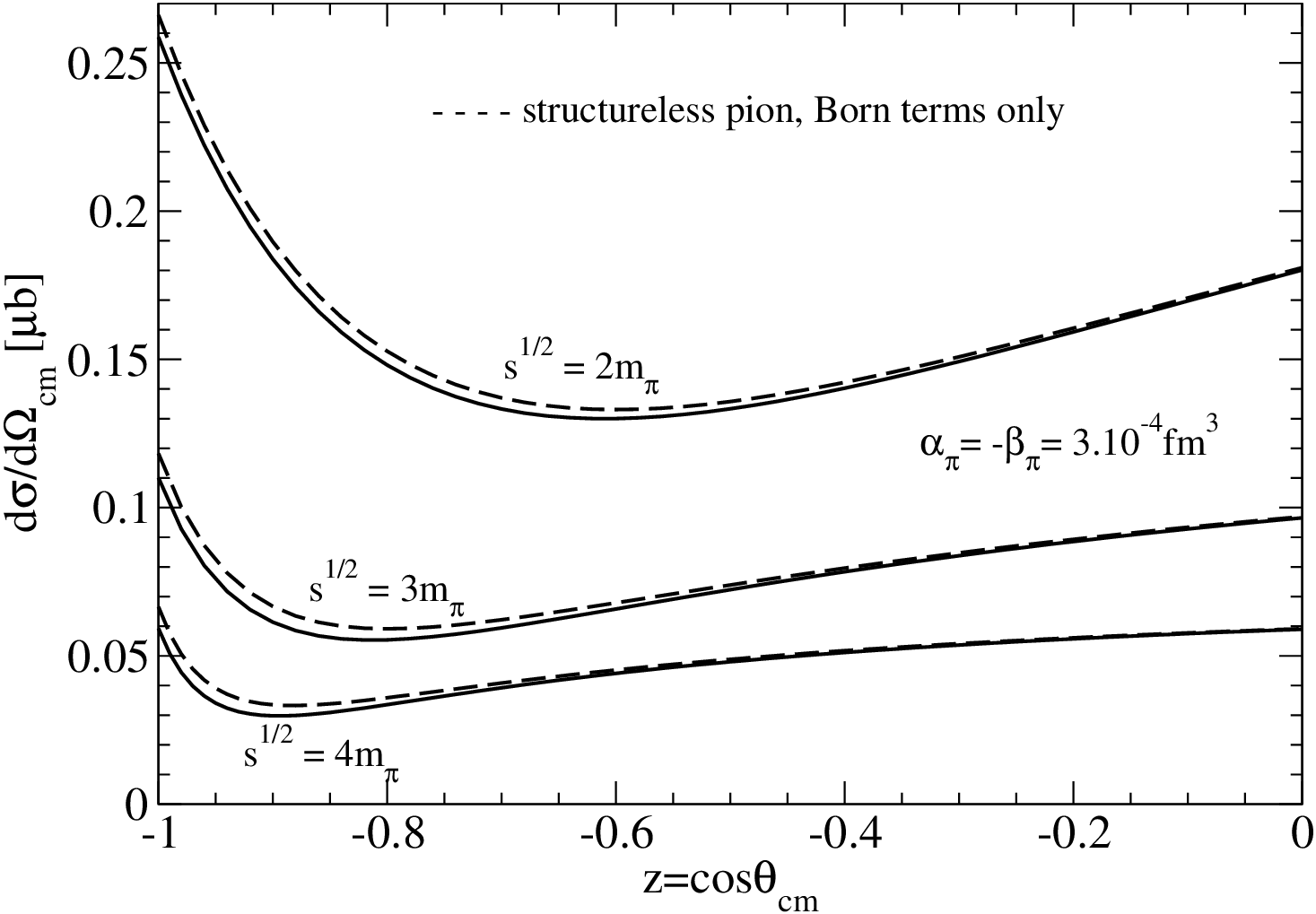}
\end{center}
\vspace{-.6cm}
\caption{Angular dependence of the differential cross section for charged pion
Compton scattering. The full curves are calculated with polarizabilities, 
$\alpha_\pi =-\beta_\pi  = 3.0 \cdot 10^{-4}\,$fm$^3$. The dashed curves represent 
the Born terms in eq.(4).} 
\end{figure}

Fig.\,1 shows the calculated total cross section $\sigma_{\rm tot}(s)$ for $\pi^-$ 
Compton scattering in the low-energy region $m_\pi < \sqrt{s} < 3.5 m_\pi$. At
higher energies the excitation of the broad $\rho(770)$-resonance becomes 
prominent ($m_\rho\simeq 5.5m_\pi$). One observes that the one-loop result is 
almost  indistinguishable from the cross section for a point-like pion: 
\begin{equation} \sigma_{\rm tot}(s)^{(\rm pt)}= {4\pi \alpha^2(s+m_\pi^2) \over s(s- 
m_\pi^2)^3 }\bigg[ s^2-m_\pi^4 -2s m_\pi^2\ln{s \over m_\pi^2}\bigg]\,.\end{equation}
The effects from the pion's low-energy structure (pion-loops and electric
minus magnetic polarizability) contribute at the level of only $1\%$. This
finding is of course not entirely new \cite{buergi}. 

The angular dependence of the differential cross section $d\sigma/d\Omega_{\rm  
cm}$ is shown in Fig.\,2 for three selected center-of-mass energies,
$\sqrt{s}=(2,3,4) \,m_\pi$. The full curves are calculated with (assumed) pion
polarizabilities $\alpha_\pi =-\beta_\pi  = 3.0 \cdot 10^{-4}\,$fm$^3$ and the
dashed lines  correspond to the case of a structureless pion:
\begin{equation} {d \sigma^{(\rm pt)}\over d\Omega_{\rm cm}}= {\alpha^2[s^2(1+z)^2+ 
m_\pi^4(1-z)^2]  \over s[s(1+z)+m_\pi^2(1-z)]^2}\,,\end{equation}
where $z = \cos \theta_{\rm cm}$. One gets instructed that the effects of the
pion's low-energy structure on observables are rather small. For $\sqrt{s} < 
4m_\pi$, they amount to at most an $11\%$ percent reduction of the differential 
cross section in backward directions, $z\approx -1$. This gives some
impression of the experimental challenge posed for measuring with good 
accuracy the pion polarizabilities. First, low-energy $\pi^-\gamma$ cross 
sections (e.g. $\sqrt{s} < 4m_\pi$) need to be extracted from the Primakoff 
events in the Coulomb peak. Secondly, these must come with an accuracy such 
that the deviations from the point-like cross section become statistically
significant. We also note that the (quadratic) pion-loop contribution
$A(s,t)^{(\rm  loop)} \sim t^2+\dots$ in eq.(5) works against the (linear) 
polarizability term $A(s,t)^{(\rm  pola)} \sim t$. Therefore, leaving it out
in an analysis of pion Compton scattering data would lead to an 
underestimation of the pion polarizabilities $\alpha_\pi \simeq -\beta_\pi>0$. 
In the optimal case their fitted value should also be independent of the upper 
limit put on the center-of-mass energy $\sqrt{s}$. 

In order to visualize more directly the effect of the pion polarizabilities, we
show in Fig.\,3 the ratio $d\sigma/d\sigma_0$ between the differential cross
sections (at one-loop order) in backward direction, $z=-1$, evaluated with 
finite polarizabilities $\alpha_\pi= - \beta_\pi=3.0 \cdot 10^{-4}\,$fm$^3$, and 
with polarizabilities set equal to zero.  At $\sqrt{s}=4m_\pi$ the effect 
becomes now quite sizeable, where it amounts to almost a $20\%$ reduction of 
the (backward) differential cross section. Fig.\,3 reemphasizes the necessity 
to include (at least) the pion-loop correction eq.(5) in the analysis of 
Primakoff data. Of course, it would also be interesting to see how the ratio
displayed in Fig.\,3 gets affected by the two-loop corrections of 
ref.\cite{gasser}, in particular because of the presence of a branch point
at $\sqrt{s}=3m_\pi$.       

\begin{figure}
\begin{center}
\includegraphics[scale=.55,clip]{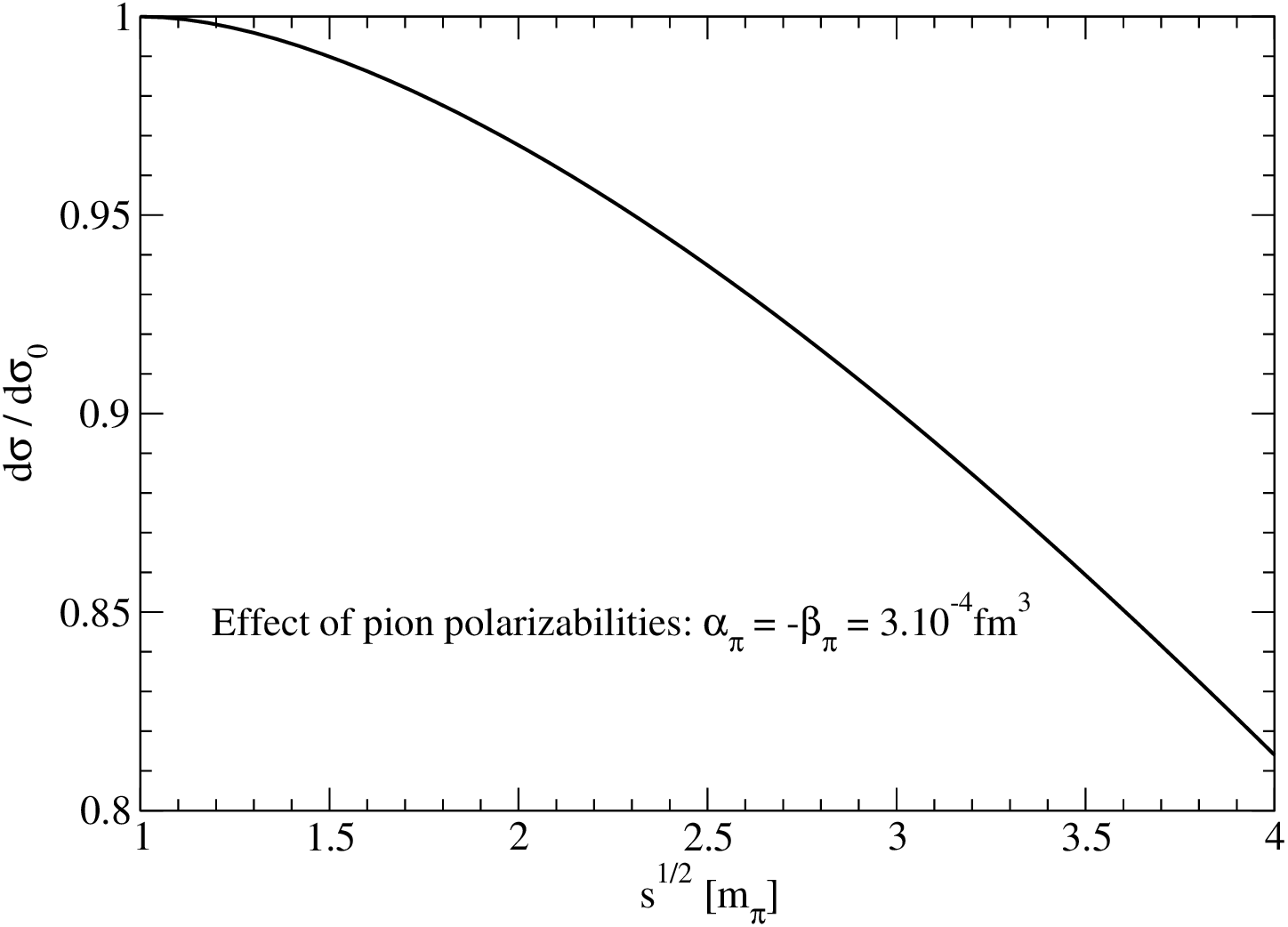}
\end{center}
\vspace{-.6cm}
\caption{Ratio $d\sigma/d\sigma_0$ between the differential cross sections in 
backward direction ($z=-1$) calculated with finite, $\alpha_\pi= - \beta_\pi=3.0
\cdot 10^{-4}\,$fm$^3$, and with zero pion polarizabilities.}
\end{figure}
      
\section{Chiral anomaly test}
Next, we come to the neutral pion production reaction $\pi^-(p_1)+ \gamma(k,
\epsilon) \to\pi^-(p_2)+\pi^0(p_0)$. The corresponding T-matrix:    
\begin{equation}T_{\gamma3\pi}={e\over 4\pi^2f_\pi^3}\,\epsilon_{\mu\nu\kappa\lambda} 
\epsilon^\mu p_1^\nu p_2^\kappa p_0^\lambda \,M(s,t)\,, \end{equation}
involves (as any process which violates natural parity) the totally 
antisymmetric $\epsilon$-tensor and $M(s,t)$ is a dimensionless invariant
function equal to $1$ in the soft-pion limit. The prefactor $F_{\gamma3\pi} =
e/(4\pi^2f_\pi^3) = 9.72\,$GeV$^{-3}$ is fixed by the chiral anomaly of QCD
(i.e. the anomalous $VAAA$ rectangle quark diagram). In the effective chiral
Lagrangian, the $\gamma3\pi $-vertex is provided by the famous
Wess-Zumino-Witten term (together with the $\pi^0\to \gamma \gamma$ decay
vertex). Neutral pion production in Primakoff reactions serves therefore as
test of the QCD chiral anomaly. The low-statistics experiment performed  some
time ago at Serpukhov has obtained the somewhat high value $F_{\gamma3 \pi}
=(12.9 \pm 1.4)\,$GeV$^{-3}$ \cite{antipov}. An improved measurement of
$F_{\gamma3\pi}$ with much higher statistics has been proposed by the COMPASS 
collaboration \cite{compass} and preliminary data are presently already 
being analyzed \cite{thimo}. Here, it should be noted that electromagnetic
corrections \cite{ametller} as well as two-loop corrections \cite{hannah}
have been calculated for the process $\pi^-\gamma \to\pi^-\pi^0$. If these
corrections are included in the analysis of the Serpukhov data the extracted 
value of $F_{\gamma3\pi}$ gets somewhat reduced to $F_{\gamma3 \pi} =(10.7 \pm
1.2) \,$GeV$^{-3}$ \cite{ametller}.

The total cross section for $\pi^-\gamma \to\pi^-\pi^0$ following from the
T-matrix in eq.(10) has the form: 
\begin{equation}\sigma_{\rm tot}(s)={\alpha(s-m_\pi^2)(s-4m_\pi^2)^{3/2}\over 
(4f_\pi)^6 \pi^4  \sqrt{s}}\int_{-1}^1\!dz \,(1-z^2)\,|M(s,t)|^2\,,\end{equation}
after substituting $2t= 3m_\pi^2-s +z(s-m_\pi^2)\sqrt{1-4m_\pi^2/s}$ with $z$ the
cosine of the $\pi^-$ cms scattering angle. The prefactor $(s-4m_\pi^2)^{3/2}$
signals that the final state pions are produced in a relative P-wave. The 
one-loop corrections for the process $\pi^-\gamma \to\pi^-\pi^0$ have been 
calculated in chiral perturbation theory some time ago in ref.\cite{bijnens}. 
Together with the tree-level contribution the invariant amplitude $M(s,t)$ 
reads: 
\begin{equation}M(s,t)^{(\rm 1-loop)}=1+ {m_\pi^2\over(4\pi f_\pi)^2}  \bigg\{{4\over 
3} \Big[ J(s) +  J(t) + J(u)\Big] -{1\over 2} -\ln{m_\pi \over m_\rho} \bigg\}
+{3m_\pi^2  \over 2m_\rho^2}\,, \end{equation}
where $s+t+u = 3m_\pi^2$ and we have introduced the loop function:  
\begin{equation}J(4m_\pi^2\, x) ={4x \over 3} -1 +(1-x) \sqrt{1-{1\over x}}\,
\bigg\{\ln|\sqrt{x}+\sqrt{x-1}|-{i \pi\over 2}\theta(x-1)\bigg\}\,.
\end{equation} 
The correction terms to $1$ in eq.(12) represent one-loop pion-rescattering in
the $s$-, $t$-, and  $u$-channel. The last term in eq.(12) corresponds to the
counterterm contribution estimated via vector meson exchange and using
simplifying symmetry relations for the coupling constants \cite{bijnens}.    

\begin{figure}
\begin{center}
\includegraphics[scale=.55,clip]{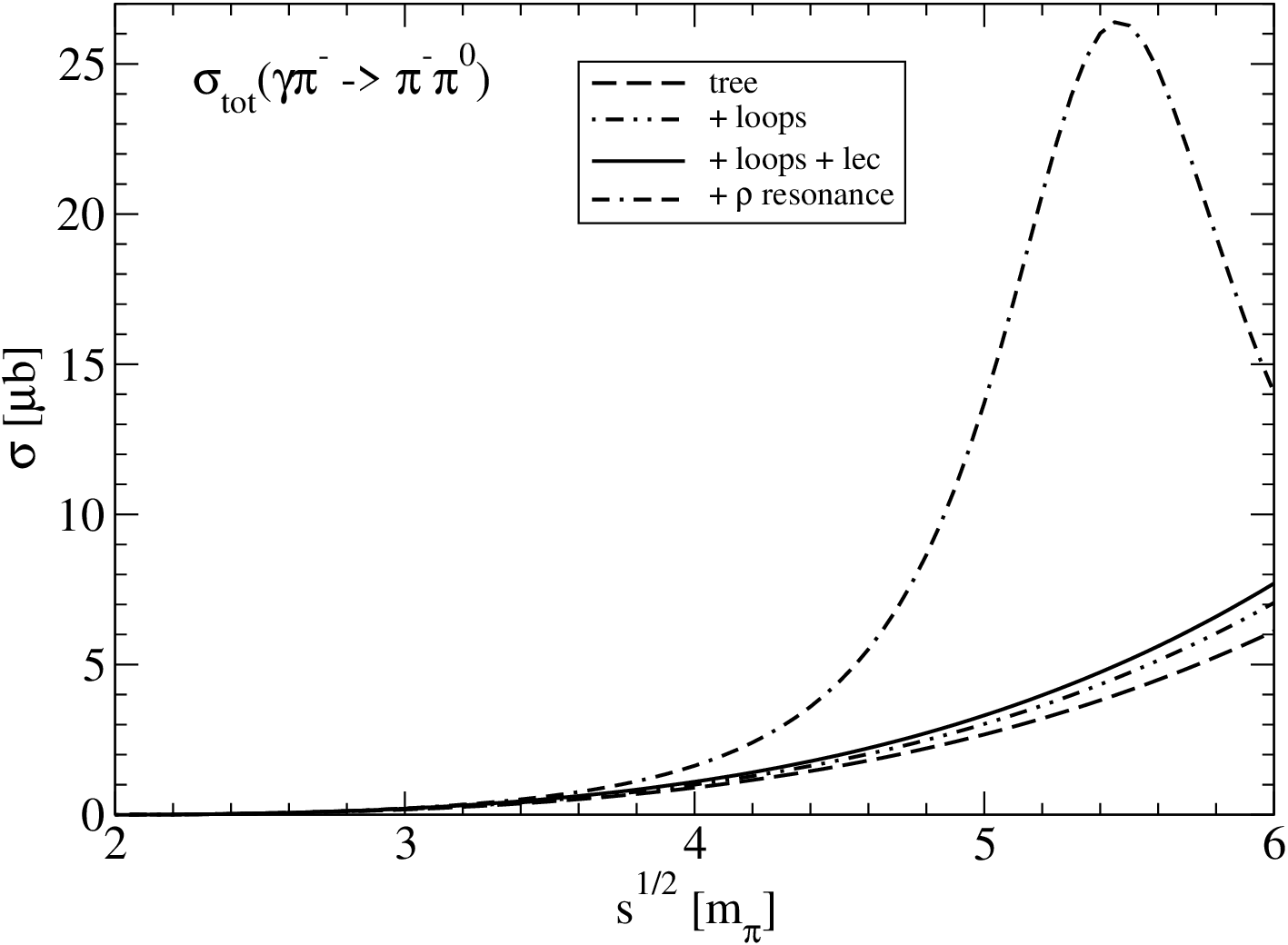}
\end{center}
\vspace{-.6cm}
\caption{Total cross sections for the reaction $\pi^- \gamma \to \pi^- \pi^0$
as a function of the center-of-mass energy $\sqrt{s}$.}
\end{figure}

The three lower curves in Fig.\,4 show the total cross section for $\pi^-
\gamma \to \pi^- \pi^0$ in the region $2m_\pi < \sqrt{s} < 6m_\pi$. One sees
that the pion-loop and the counterterm corrections each enhance the total cross
section by about $10\%$ in comparison to the tree-level approximation. Above 
center-of-mass energies of $\sqrt{s} \simeq 4m_\pi$ the effects of the 
prominent  $\rho(770)$-resonance can no more be represented by a local
counterterm. This is clearly demonstrated by the upper dashed-dotted curve in
Fig.\,4 which stems from  a $\rho$-meson exchange model:         
\begin{equation} M(s,t)^{(\rho)} = 1 + {2g_{\rho\pi} G_{\rho \pi \gamma}\over m_\rho^3  
F_{\gamma3\pi}}\bigg\{{s\over m_\rho^2 -s -i\sqrt{s}\,\Gamma_\rho(s)} +{t
\over m_\rho^2-t} +{u \over  m_\rho^2 -u} \bigg\}\,, \end{equation}
including in the $s$-channel exchange term an energy-dependent $\rho$-meson
decay  width:   
\begin{equation}\Gamma_\rho(s) ={g_{\rho\pi}^2\over   48\pi s}(s-4m_\pi^2)^{3/2} 
\,. \end{equation} 
Here, $g_{\rho \pi} = 6.03$ is determined from the empirical decay width 
$\Gamma_\rho(m_\rho^2)=150\,$MeV \cite{pdg}, and the coupling constant $G_{\rho
  \pi  \gamma}\simeq 0.17$ can be inferred from the empirical branching ratio
Br$(\rho \to \pi \gamma) = (4.5\pm 0.5) \cdot 10^{-4}$ \cite{pdg}. At the
$\rho$-meson peak the resonant  cross section exceeds the one due to the chiral
low-energy terms by about a factor 5. Potentially, this spectral shape offers 
a strategy to avoid absolute cross section measurements, namely by covering in 
the COMPASS experiment the whole region from the $\rho$-peak down to 
threshold. Yet, some model dependence will remain in  such a procedure of 
extracting  $F_{\gamma3\pi}$. Clearly, a careful analysis of the near-threshold 
data for $\pi^-\gamma \to \pi^- \pi^0$ should the include radiative and higher 
loop corrections of refs.\cite{ametller,hannah}.

\section{Double pion production} 
Next, we turn to double neutral pion production $\pi^-\gamma \to\pi^-\pi^0
\pi^0$. This reaction has not been considered so far, mainly because of the
lack of any experimental data. With the expected high-statistics of the
COMPASS experiment \cite{compass} this may change in the near future.   

\begin{figure}
\begin{center}
\includegraphics[scale=1.,clip]{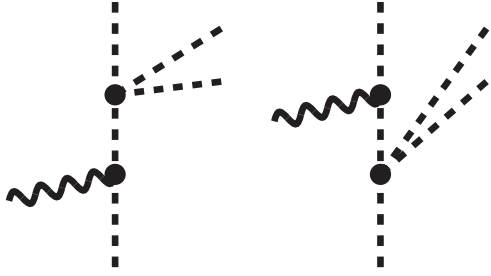}
\end{center}
\vspace{-.5cm}
\caption{Tree level diagrams for double neutral pion production: $\pi^- \gamma 
\to \pi^- \pi^0\pi^0$. In the case of charged pion-pair production, $\pi^-\gamma 
\to \pi^- \pi^+\pi^-$, the incoming photon couples also to the other two
outgoing  pions.}
\end{figure}

The tree level diagrams for $\pi^-\gamma \to\pi^-\pi^0\pi^0$ are shown in 
Fig.\,5. In the convenient parametrization of the special-unitary matrix field 
$U=\sqrt{1-\vec \pi^{\,2}/f_\pi^2}+i \vec \tau \cdot \vec \pi/f_\pi$ a 
$\gamma 4\pi$ contact-vertex does not exist. (In other parametrizations its 
contribution gets canceled by the off-shell part of the $4\pi$-vertex.) 
Moreover, since the left diagram vanishes  in Coulomb gauge $\epsilon\cdot
(2p+k)=0$ we need to evaluate only one single diagram. Performing the 
polarization sum and three-body phase space integration, we end up with the 
following expression for the total cross section for $\pi^-\gamma \to 
\pi^-\pi^0 \pi^0 $:
\begin{eqnarray}\sigma_{\rm tot}(s)&=&{\alpha\over 32\pi^2 f_\pi^4 (s-m_\pi^2)^3} 
\int_{2m_\pi \sqrt{s}}^{s-3m_\pi^2}\!\!dw \,\sqrt{s-w-3m_\pi^2\over s-w  +m_\pi^2}\,(s-w)^2 
\nonumber \\ && \times\Bigg\{w \ln{w+\sqrt {w^2-4m_\pi^2 s}\over 2m_\pi  \sqrt{s}}
-\sqrt{w^2-4m_\pi^2s} \Bigg\}\,.  \end{eqnarray}
Here, the integration variable $w$ is $2\sqrt{s}$ times the cms energy of the 
outgoing $\pi^-$, and the factor $(s-w)/f_\pi^2$ originates from the chiral 
$\pi^+\pi^- \to \pi^0\pi^0$ interaction.

The lower dashed curve in Fig.\,6 shows this total cross section as a function
of the center-of-mass energy in the region $3m_\pi < \sqrt{s}<6m_\pi$. Near
threshold it grows cubically with the excess energy, $\sigma_{\rm tot}(s)_{\rm
thr}=\alpha \sqrt{3}\, (\sqrt{s}-3m_\pi)^3/[\pi m_\pi (8f_\pi)^4]$. With values up 
to one microbarn, this cross section is still quite sizeable. Note that a 
resonant $\rho$-meson contribution is now not possible, since $\rho\to 3\pi$ 
is forbidden by G-parity.    

For the sake of completeness we treat also the charged pion-pair production 
process $\pi^-\gamma\to \pi^-\pi^+\pi^- $. Two additional diagrams where the
photon couples to the other outgoing pions, need then to be included. Putting 
all pieces together, we end up with the following expression for total cross
section for $\pi^-\gamma\to \pi^-\pi^+\pi^-$: 
\begin{eqnarray} \sigma_{\rm tot}(s)&=&{\alpha\, s\over 8\pi^3 f_\pi^4 (s-m_\pi^2)^3} 
\int\limits_{z^2<1}\!\!\!\!\!\int\!d\omega_1 d\omega_2 \int_{-1}^1 
\!dx\int_0^\pi \! d\phi \,\Bigg\{ {q_1^2(1-x^2) \over (\omega_1-q_1 x)^2}
\nonumber \\ &&\times\bigg\{2\big[p_0(\sqrt{s}-\omega_2)-\sqrt{s}\omega_1-k_0q_2 
y\big]^2 +\big[k_0 \sqrt{s}-p_0\omega_1-k_0 q_1 x\big]^2 \bigg\} \nonumber \\
&&+{2q_1q_2(z-x y) \over (\omega_1 -q_1 x)(\omega_2-q_2 y)}\bigg\{\big[p_0
(\sqrt{s}-\omega_2)-\sqrt{s}\omega_1-k_0 q_2 y\big] \nonumber \\ &&\times 
\big[p_0(\sqrt{s}-\omega_1)-\sqrt{s}\omega_2-k_0 q_1 x\big]+ 2\big[k_0\sqrt{s} 
-p_0\omega_2-k_0 q_2 y\big] \nonumber \\ && \times \big[p_0(\omega_1+\omega_2)-
\sqrt{s}\omega_1+k_0 ( q_1 x+q_2 y)\big] \bigg\} \Bigg\} \,, \end{eqnarray} 
with $p_0 = (s+m_\pi^2)/(2\sqrt{s})$,  $k_0 = (s-m_\pi^2)/(2\sqrt{s})$, $q_{1,2} = 
\sqrt{\omega_{1,2}^2-m_\pi^2}$ and 
\begin{equation} q_1q_2 \,z =\omega_1 \omega_2-\sqrt{s}(\omega_1 +\omega_2) 
+{s+m_\pi^2 \over 2}  \,, \quad\quad y = xz +\sqrt{(1-x^2)(1-z^2)} \cos\phi \,.
\end{equation} 

\begin{figure}
\begin{center}
\includegraphics[scale=.55,clip]{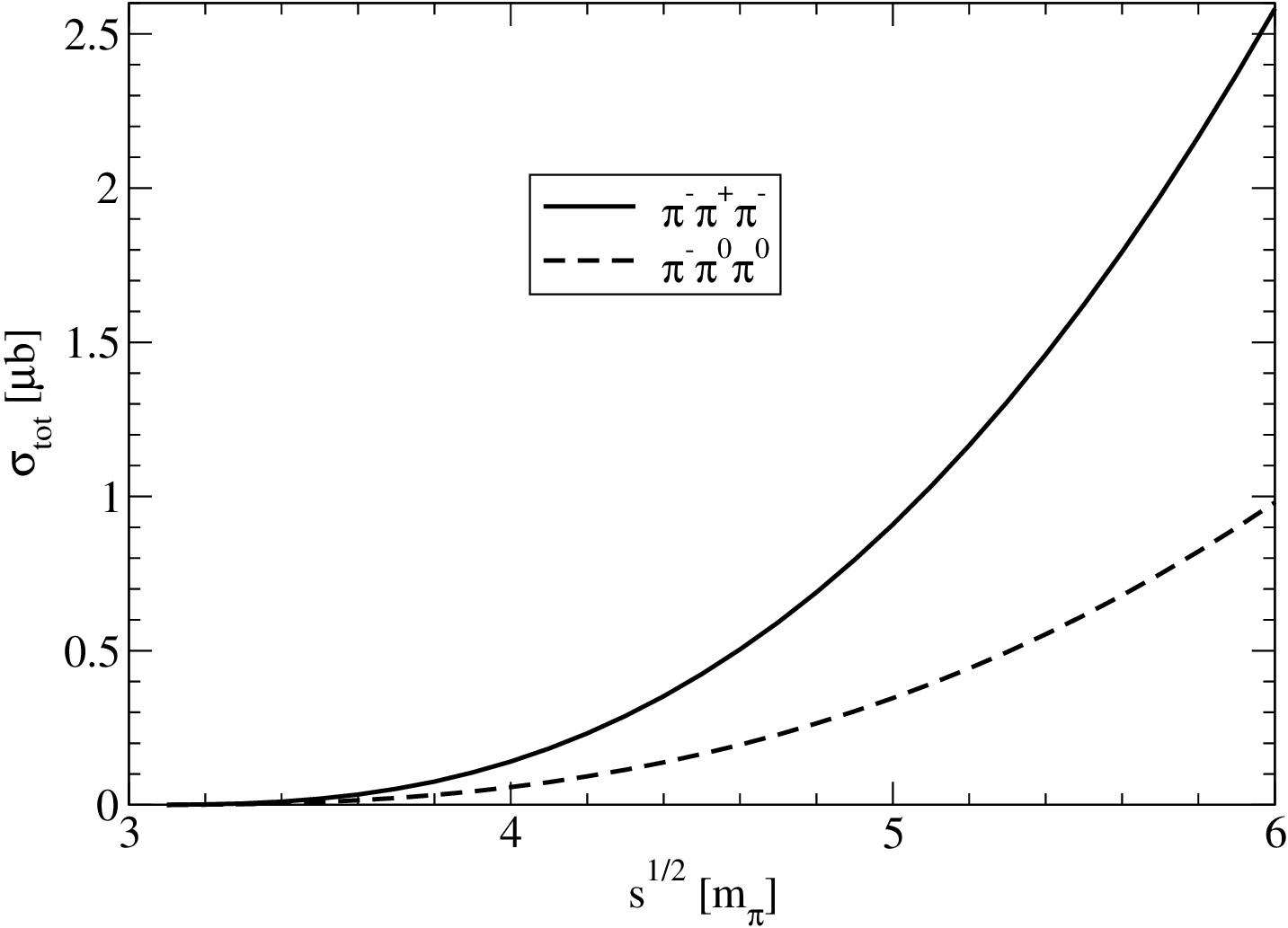}
\end{center}
\vspace{-.4cm}
\caption{Total cross sections for the double pion production reactions $\pi^- 
\gamma \to \pi^- \pi^0 \pi^0$ and $\pi^- \pi^+\pi^-$ as a function of the total 
center-of-mass energy $\sqrt{s}$.}
\vspace{-.3cm}
\end{figure}
In eq.(17) we have exploited the permutational symmetry of the three-pion
phase space in order reduce the number of independent interference terms in
the integrand. This cross section grows in the threshold region also cubically 
with the excess energy, $\sigma_{\rm tot}(s)_{\rm thr}=\alpha \sqrt{3}\, (\sqrt{s}-
3m_\pi)^3/[9\pi m_\pi (4f_\pi)^4]$.  

After numerical evaluation of the four-dimensional integral we obtain the total
cross sections shown by the upper full curve in Fig.\,6. They come out about a 
factor $2.5$ larger than the ones for $\pi^- \gamma \to \pi^- \pi^0 \pi^0$,
reaching $2.6\,\mu$b at $\sqrt{s} = 6m_\pi$. This enhancement can be traced
back to the larger number of contributing diagrams. At their upper ends the 
curves in Fig.\,6 should be considered only as indicative since there higher 
orders may become significant.

\section{Radiative correction: $\pi^-\gamma\to \pi^-\gamma\gamma $} 
Finally, we consider the radiative correction process $\pi^-\gamma\to
\pi^-\gamma \gamma$ to pion Compton scattering. As a smooth background process 
it interferes also with the neutral pion production $\pi^-\gamma\to \pi^-
\pi^0$. Therefore a knowledge of its order of magnitude will be helpful in
analyzing the Primakoff data. 
\begin{figure}
\begin{center}
\includegraphics[scale=0.95,clip]{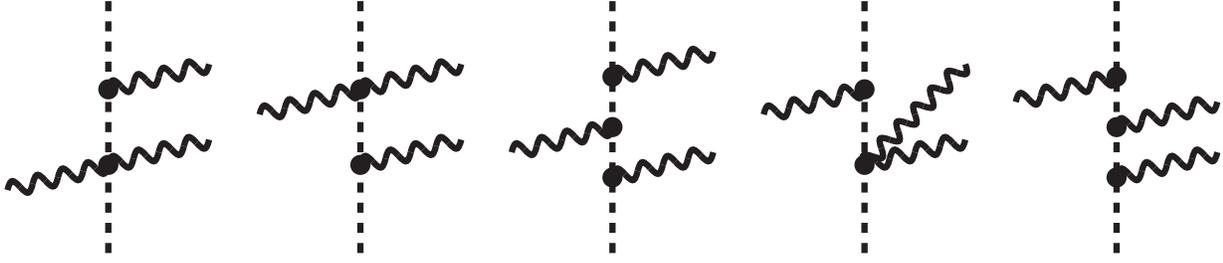}
\end{center}
\vspace{-.4cm}
\caption{Tree level diagrams for $\pi^- \gamma \to \pi^- \gamma\gamma$. The
coupling of the incoming photon at the bottom of the pion line vanishes in
Coulomb gauge. Diagrams with the two outgoing photons interchanged are not
shown.}
\vspace{-.3cm}
\end{figure}
A representative set of tree diagrams is shown in Fig.\,7. These need to be 
supplemented by diagrams with the two outgoing (right) photons interchanged. 
We exploit the fact that the coupling of the incoming (left) photon at the
bottom of the pion line  vanishes in Coulomb gauge and thus are left with 9
non-vanishing diagrams. After performing the triple sums over transversal
photon polarizations  for the 45 interference terms and applying flux and
phase space factors, we end up with a double differential cross section of the
form:    
\begin{equation}{ d^2\sigma\over d\omega_1 d\omega_2}={\alpha^3\over8\pi(s-
m_\pi^2)} \int_{-1}^1 \!dx\int_0^\pi \!d\phi\,R(\sqrt{s},m_\pi^2,\omega_1, 
\omega_2,x,y,z)\,. \end{equation} 
Here, $R(\dots)$ stands for a rational function of its six arguments (which is
much too long to be reproduced here).  $\omega_1$ and $\omega_2$ denote the cms
energies of the final state photons and
\begin{equation} z = 1-\sqrt{s}\bigg({1\over \omega_1}+  {1\over \omega_2}\bigg)
+{s-m_\pi^2 \over 2 \omega_1 \omega_2}\,, \quad\quad y = xz +\sqrt{(1-x^2)(1-z^2)}
\cos\phi \,. \end{equation}
\begin{figure}
\begin{center}
\includegraphics[scale=.55,clip]{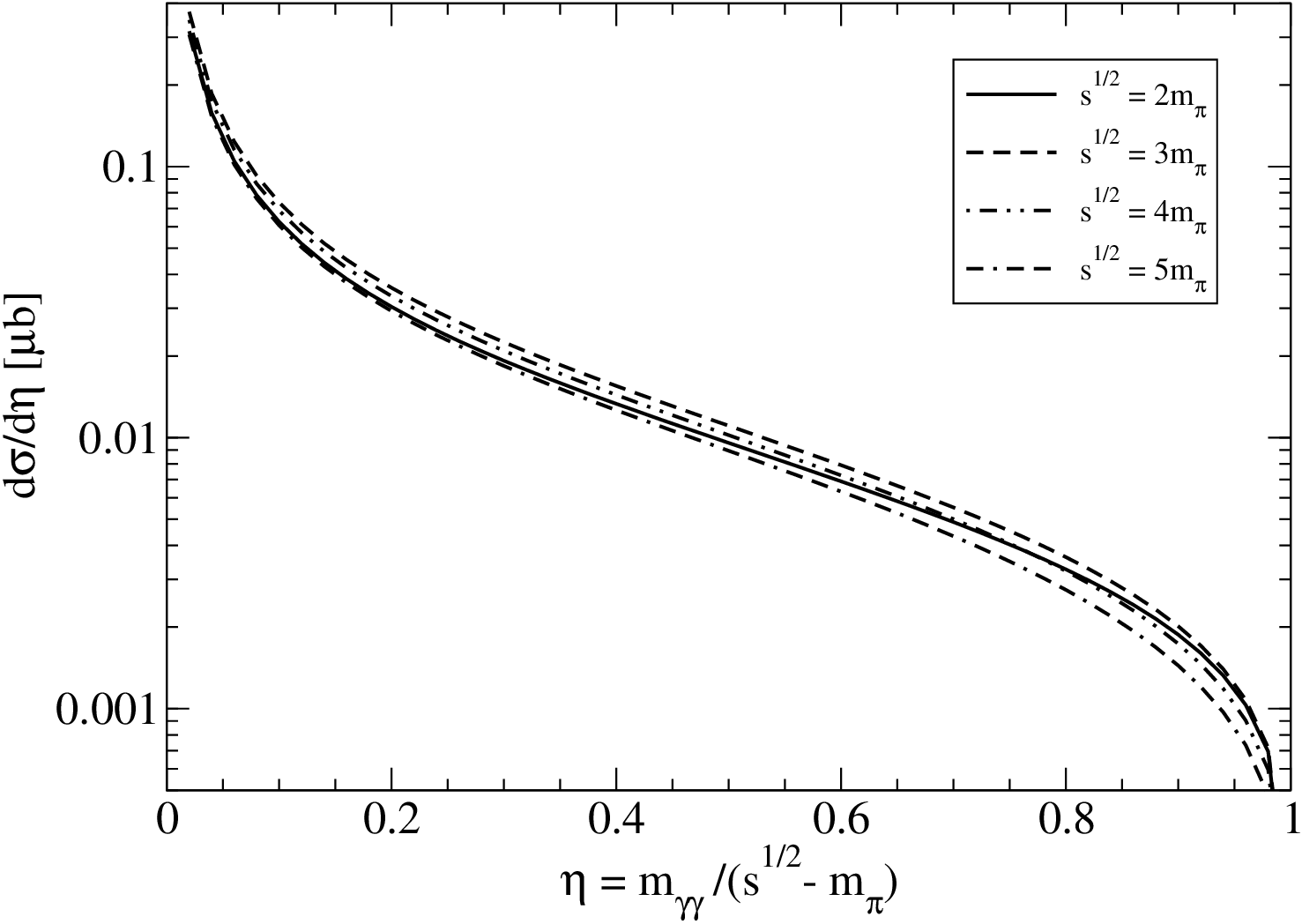}
\end{center}
\vspace{-.8cm}
\caption{Two-photon mass spectra for the reaction $\pi^- \gamma \to \pi^-
\gamma\gamma$ as a function of the variable $\eta=
m_{\gamma\gamma}/(\sqrt{s}-m_\pi)$.}
\end{figure}
The two equations $z=1$ and $z=-1$ determine the phase space boundaries as a
straight line and a hyperbola in the $\omega_1\omega_2$-plane. It is well-known
that soft photons with $\omega_{1,2}\to 0$ cause a logarithmic infrared 
divergence of the integrated (total) cross section. Therefore, we introduce as 
a suitable Lorentz-invariant variable the $\gamma\gamma$ invariant mass 
squared:
\begin{equation} m_{\gamma\gamma}^2 = 2 \sqrt{s}(\omega_1 +\omega_2)   +m_\pi^2-s = 
(\sqrt{s}-m_\pi)^2 \cdot \eta^2 \,, \end{equation}  
related to the sum of the photon energies $\omega_1 +\omega_2$ and integrate
over their half difference $(\omega_1 -\omega_2)/2$. For the purpose of a 
uniform representation we express the two-photon mass spectrum $d \sigma/d
\eta$ in terms of the dimensionless variable $\eta$ which ranges for all
$\sqrt{s}$ in the unit interval, $0 \leq  \eta \leq 1$. 

Fig.\,8 shows the calculated two-photon mass spectrum $d\sigma/d\eta$ for the 
process $\pi^- \gamma \to \pi^- \gamma\gamma$ at four values of $\sqrt{s}=(2,3,
4,5)\,m_\pi$. For $\eta\to 0$ the curves diverge as $d\sigma/d\eta \sim 
\eta^{-1}$. It is astonishing that the $\gamma\gamma$-spectra possess only a 
very weak dependence on the total center-of-mass energy $\sqrt{s}$. A possible
reason herefore could be the similar weak variation of the ratio $\pi
\gamma\gamma$ three-body phase space over flux factor: $1+m_\pi^2 s^{-1}
-2m_\pi^2 (s-m_\pi^2)^{-1} \ln(s/m_\pi^2)$.  

We have also considered the interference term of the bremsstrahlung diagrams
in Fig.\,7 with the $2\gamma$-production from $\pi^0$-decay. The amplitude for
the latter process involves the large scale factor $2\sqrt{2} \pi f_\pi =
821\,$MeV to the fourth power in the denominator. As a consequence of that
suppression factor, the integrated interference cross sections (taking
e.g. $0.01 \leq \eta \leq 1)$ turn out to be extremely small.     

\begin{figure}
\begin{center}
\includegraphics[scale=.55,clip]{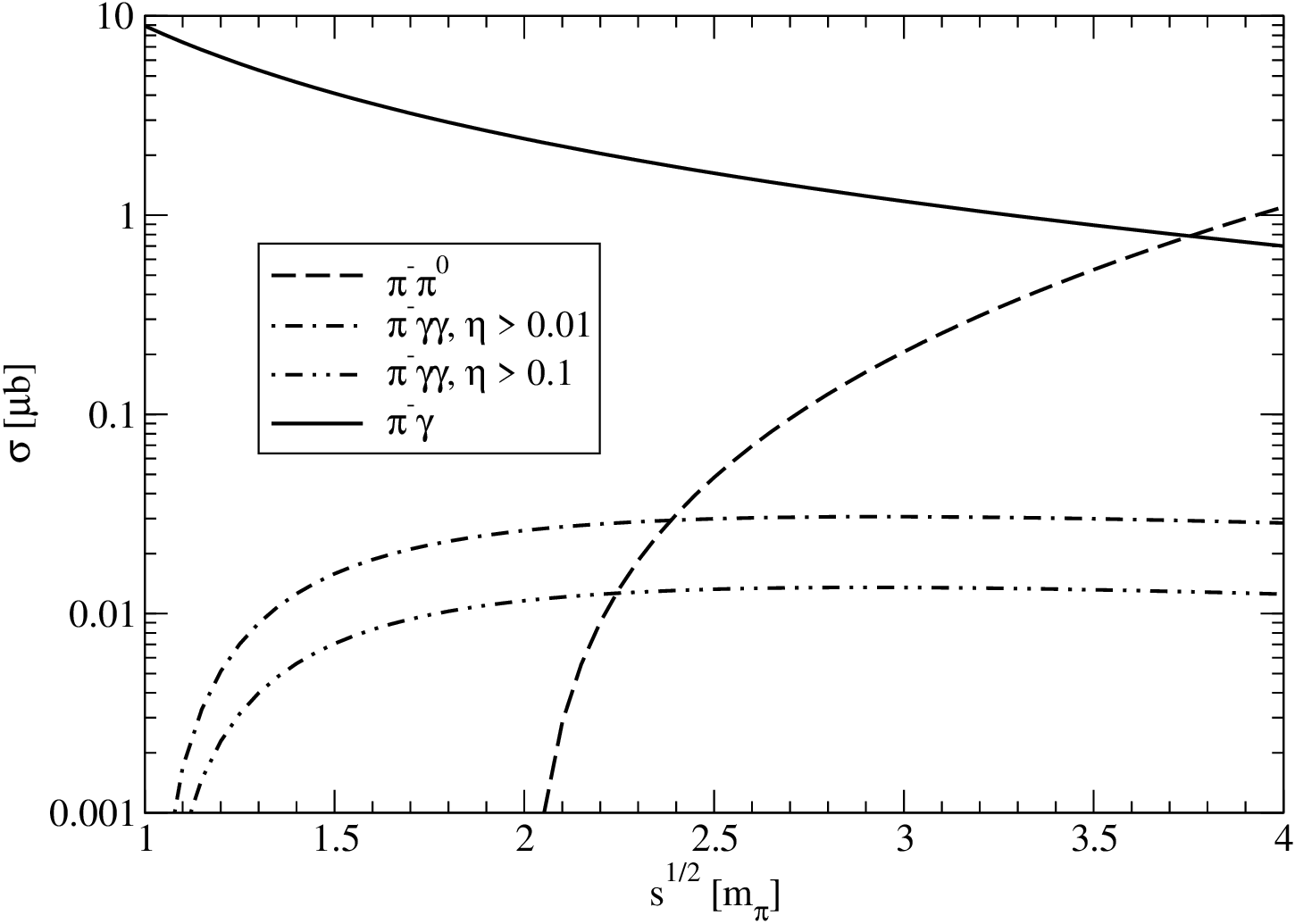}
\end{center}
\vspace{-.8cm}
\caption{Comparison of total cross sections for the final states $\pi^-\gamma$, 
$\pi^-\pi^0$, and $\pi^-\gamma\gamma$. The latter ones have been obtained as
infrared-regularized integrals $\int_{\eta_{\rm min}}^1d\eta \,(d\sigma/d\eta)$ with 
$\eta_{\rm  min}=0.01$ and  $0.1$.}
\end{figure}

Moreover, we show in Fig.\,9 (on a logarithmic scale) the total cross sections
for the three final states $\pi^- \gamma$ (full curve), $\pi^-\pi^0$ (dashed
curve), and $\pi^- \gamma \gamma$ (dashed-dotted curves), altogether. In the
latter case these have been obtained (in infrared-regularized form) by 
integrating the two-photon mass spectrum $d\sigma/d\eta$ over the interval 
$0.01 \leq \eta \leq 1$ and $0.1\leq \eta \leq 1$, respectively. With the 
exception of the region $ \sqrt{s}<2.5m_\pi$ very near to the 
$\pi^0$-production threshold, the non-resonant two-photon radiation amounts to 
at most a few percent correction. 

As already mentioned, the two-photon mass spectrum $d\sigma/d \eta$ of the
reaction $\pi^- \gamma \to \pi^- \gamma \gamma$  possesses an infrared 
singularity of the form: 
\begin{equation} { d \sigma \over d\eta} = {\sigma_{\rm ir}(s) \over \eta}+ {\cal
O}(\eta) \,, \end{equation}
in the limit $m_{\gamma\gamma}\to 0$. In practice the leading $1/\eta$ term 
provides a good description of the two-photon mass spectrum up to $\eta 
\simeq 0.25$. After an elaborate calculation we obtain the following 
analytical expression for the energy dependent function  $\sigma_{\rm ir}(s)$ 
associated with the infrared singular part:
\begin{eqnarray}\sigma_{\rm ir}(s) &=& {32 \alpha^3 \over 3s} \bigg\{-2 +{\hat s+1
\over \hat s-1}\ln\hat s \bigg\} \nonumber \\ && +{16\alpha^3 \over 3s}\bigg\{  
{\hat s^2+10\hat s+1 \over (\hat s-1)^2}-{6\hat s(\hat s+1)\over(\hat s-1)^3} 
\ln\hat s \bigg\} \bigg\{ {2\hat s-7\hat s^2-7 \over 6(\hat s-1)^2}+{\hat s(
\hat s+1)\over(\hat s-1)^3}\ln\hat s \bigg\}\nonumber \\ &&+{64\alpha^3\over s}
\int_0^1 \!\!dx \, {(1-x)\sqrt{x}\big[(1-x)^2\hat s-x^2\big]\big[x (\hat s -1)^2 
+2 \hat s\big] \over  (\hat s+x-x \hat s)^2 \sqrt{x(\hat s-1)^2+4\hat s }}
\nonumber \\ && \qquad \qquad \qquad \times \ln {\sqrt{x}(\hat  s -1)+\sqrt{x(
\hat s -1)^2+4\hat s } \over 2 \sqrt{\hat s}} \,, \end{eqnarray}
where $\hat s = s/m_\pi^2$. The corresponding
behavior near threshold is, $\sigma_{\rm ir}(s)= 64\alpha^3 (\sqrt{s}-m_\pi)^2/
(9m_\pi^4)$, and it comes exclusively from the first term in eq.(23). In order 
to arrive at this result we have first established in a careful numerical 
study the 20 non-vanishing contributions to $\sigma_{\rm ir}(s)$ from the 
interference terms of the diagrams in Fig.\,7 as well as the identities 
holding between them.  As a result, only six independent interference terms 
needed to be considered and for these the three-dimensional phase space 
integral over $((\omega_1-\omega_2)/2,x,\phi$) could then be solved (almost) 
analytically. In order to bring the integral-term in eq.(23) into its compact
form we substituted $x\to 1-2x$ at the end. Fig.\,10 shows the cross section 
$\sigma_{\rm ir}(s)$ as a function of the total center-of-mass energy $\sqrt{s}$. 
Quite interestingly, it reaches a maximum of $7.45\,$nb at $\sqrt s = 3m_\pi$ 
and beyond that it decreases slowly as $s^{-1} \ln(s/m_\pi^2)$. This behavior is 
in accordance with the weak $\sqrt{s}$-dependence of the two-photon mass
spectra $d\sigma/d\eta$ observed in Fig.\,8. 
     
\begin{figure}
\begin{center}
\includegraphics[scale=.55,clip]{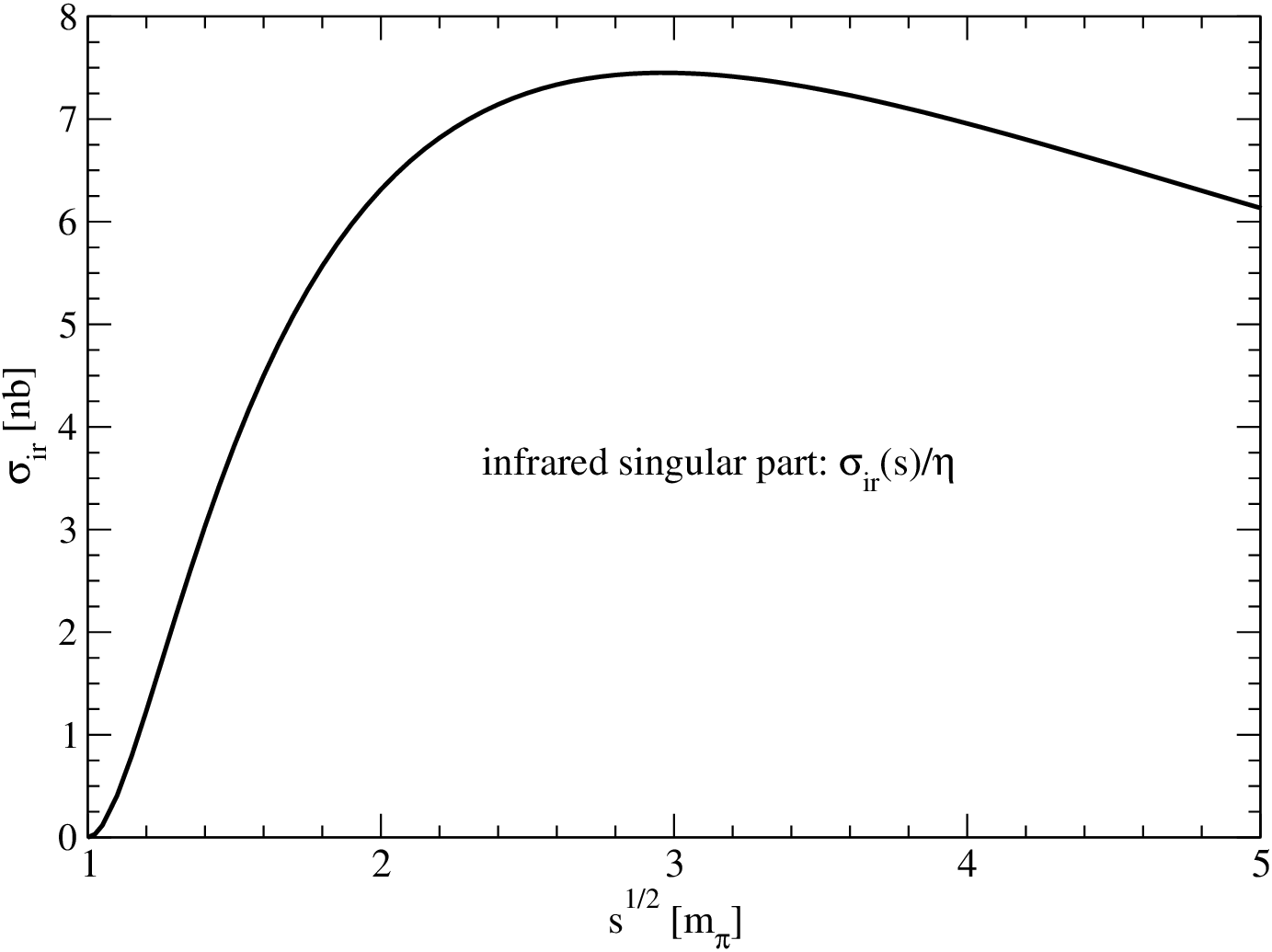}
\end{center}
\vspace{-.8cm}
\caption{Coefficient $\sigma_{\rm ir}(s)$ of the infrared singular part of the
two-photon mass spectrum.}
\end{figure}

Clearly, in order to establish the complete infrared finiteness of the QED 
radiative corrections to pion Compton scattering (\`a la Bloch-Nordsieck), one 
has to evaluate and include the corresponding one-photon loop diagrams. Work
along this line is in progress \cite{prog}. 
     
\section*{Acknowledgments}
We thank S. Paul for triggering this work and for informative discussions. We 
thank J. Gasser for critical reading of the manuscript and for suggesting
improvements. 

\end{document}